\documentclass[aps,pra,epsf,twocolumn,superscriptaddress,showpacs,preprintnumbers,amsmath,amssymb,floatfix,8pt]{revtex4-1}
\usepackage{amssymb,amsmath}
\usepackage{graphicx}
\usepackage{subfigure}
\usepackage[english]{babel}
\usepackage{float}
\usepackage{color}

\newcommand{\be}{\begin{equation}}
\newcommand{\ee}{\end{equation}}

\begin{document}

\title{Linear impurity modes in an electrical lattice: Theory and Experiment}

\author{M. I. Molina}
\affiliation{Departamento de F\'{\i}sica, 
    Facultad de Ciencias, Universidad de Chile, Santiago, Chile.}
  \email{mmolina@uchile.cl}
\author{L. Q. English}
\affiliation{Department of Physics and Astronomy, Dickinson
  College, Carlisle, Pennsylvania 17013, USA}
\author{Ming-Hua Chang}
\affiliation{Department of Physics and Astronomy, Dickinson
  College, Carlisle, Pennsylvania 17013, USA}
\author{P. G. Kevrekidis}
\affiliation{Department of Mathematics and Statistics, University of
Massachusetts, Amherst, Massachusetts 01003-4515, USA}
\affiliation{Mathematical Institute, University of Oxford, Oxford, OX2
  6GG, UK}

\begin{abstract}
We examine theoretically and experimentally the localized 
electrical modes existing in a bi-inductive electrical lattice
containing a bulk or a surface capacitive impurity. By means of the
formalism of lattice Green's functions, we are able to obtain closed-form 
expressions for the frequencies of the impurity (bound-state) eigenmodes and
for their associated spatial profiles. This affords us a
systematic understanding of how these mode properties change as a function
of the system parameters. We test these analytical results 
against experimental measurements, in both the bulk and
surface cases, and find very good agreement. Lastly, we turn to a series of quench experiments,
where either a parameter of the lattice or the lattice geometry itself is rapidly switched between two values or configurations.
In all cases, we are able to naturally explain
the results of such quench experiments from the larger analytical
picture that emerges
as a result of the detailed characterization of the impurity-mode solution branches. 
\end{abstract}

\maketitle

\section{Introduction}

In solid-state physics, it is a well established fact that
lattices can feature localized modes in the presence of {point defects, such as chemical impurities, vacancies or interstitials,}
that break the shift translational invariance of the perfect
lattice~\cite{Maradudin,Lif}.
This type of disorder-induced localization {can change the macroscopic properties of the crystal, its photon absorption spectrum (i.e., color) being a well-known example, but can} also be
responsible for a variety of interesting scattering
phenomena~\cite{marad}.
Related applications abound in a diverse array of
themes, ranging from defect modes in photonic crystals \cite{Joann} as
well as optical waveguide arrays~\cite{Pes_et99,Mor_et02,kipp} to
superconductors~\cite{andreev}, and from
dielectric super-lattices with embedded
defect layers~\cite{soukoulis} to electron-phonon
interactions~\cite{tsironis} and granular crystals in materials
science~\cite{kavou}. 

A related theme that is of particular interest concerns
the existence of localized modes at the surface (i.e., at the ends)
of linear chains due to the finite size effect or, otherwise stated,
the imposition of different kinds of boundary conditions. This
type of exploration has also been a topic of consideration
for over half a century, dating back to the early considerations
of R.F. Wallis on the effects of free ends in
one-dimensional~\cite{wallis1d},
as well as surface modes in two- and three-dimensional
lattices~\cite{wallis2d3d}. It is remarkable that the relevant
considerations continue to inspire recent work, either on the
theoretical role of different types of boundary conditions~\cite{luo},
or on modern applications including those of phononic
crystals~\cite{jk}
that may also involve transversal-rotational modes~\cite{tournas}. 

In the present work, we take advantage of the well established
framework of electrical transmission lines~\cite{remoissenet1,
  remoissenet2} as a prototypical setting where the theory of linear
impurity modes can be tested. Upon modeling the experimental setting
of recent experiments such as~\cite{english1,english2,english3} at the
linear level (i.e., in the absence of nonlinearity), we utilize the
formulation
of Green's functions~\cite{green1,green2,green3} in order
to identify both the eigenvalues/eigenfrequencies and eigenfunctions
of the linear modes of the lattice, focusing naturally on the
localized vibrations thereof. We provide analytical expressions for
these and a systematic comparison with the corresponding experimental
results. This is done not only for the linear lattice with a defect,
but
also in the case of existence of a localized surface mode.
Very good  agreement is achieved between the
two. Nevertheless,
in addition to that, there are some interesting observations and
twists offered. In particular, it is found that in each range of
frequencies (i.e., both above and below the linear band), only one
mode
can be identified theoretically and observed
experimentally. Additionally, quench type experiments
are performed where we vary the parameter controlling the detuning
from the homogeneous limit. In these we explore how the experimental
lattice ``jumps'' from one value to the other and the lattice response
accordingly transforms itself under such a quench. {Finally, we also examine the voltage dynamics on the lattice during a switching transformation in its geometry from a line with an edge impurity into a ring  with a bulk impurity.}

Our presentation will be structured as follows. In section II, we will
present the physical setup and mathematical model. In section III,
we will provide details of the Green's function formalism for
identifying the
corresponding localized modes. In section IV, we present our
corresponding
experimental results. Finally, in setion V, we summarize our
conclusions
and provide some directions for future study.

\section{The model}

\begin{figure}[t]
\includegraphics[angle=0, width=1.0\linewidth]{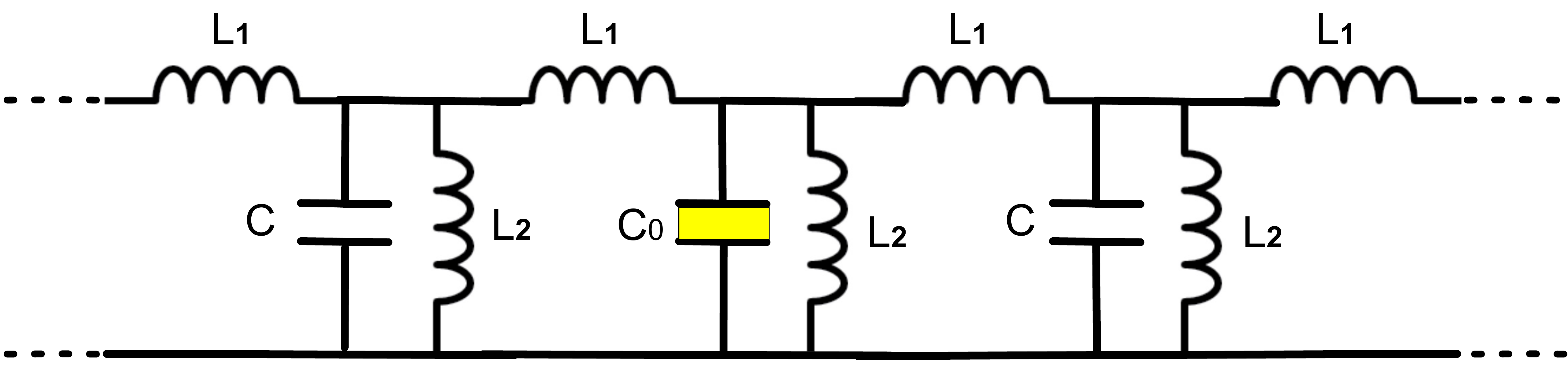}
\caption{\protect
Bi-inductive electrical lattice containing a single capacitive
impurity (in line with Remoissenet et.al.~\cite{remoissenet1, remoissenet2}).}
\label{fig1}
\end{figure}

Our physical setup is that of~\cite{remoissenet1, remoissenet2} (see
also~\cite{english1,english2,english3} for some recent experimental
studies in this system). In particular, we
consider the bi-inductive electrical lattice shown in
Fig.~\ref{fig1}.
It consists on $N$ units composed of LC resonators $\{L_{2}, C_{n}\}$
coupled inductively by an
inductor $L_{1}$.
If we call $Q_{n}$ and  $V_{n}$, respectively
the charge stored on the nth capacitor and the voltage drop  across
the inductor $L_{2}$,
the use of the Kirchhoff laws leads to a system of coupled equations
\be
{d^{2} Q_{n}\over{d t^2}} = {1\over{L_{1}}} (\ V_{n+1}+V_{n-1}-2 V_{n}
\ ) - {1\over{L_{2}}} V_{n}.\label{eq1}
\ee
The impurity at the center of our current considerations is assumed to be
a capacitive one located at $n=0$, with capacitance $C_{0}$. The rest of the capacitances are taken as identical and equal to $C$. In other words, $C_{n} = C + (C_{0} - C) \delta_{n 0}$.

By taking $Q_{n}= C_{n} V_{n}$ and $V_{n}(t) \sim \cos(\Omega t +
\phi_{n})$, we arrive at the stationary equations, i.e., the
eigenvalue problem:
\begin{eqnarray}
\Omega^2\ U_{n}&=& -\omega_{1}^2 \ (\ U_{n+1}+U_{n-1}-2 U_{n}\ ) +
                   \omega_{2}^2 U_{n}  \nonumber \\
 & & - (\Delta C/C)\ \Omega^2 \delta_{n0}\  U_{n},
\label{eqn2}
\end{eqnarray}
where $\Delta C = C_{0}-C$ and $U_{n}=V_{n }/V_{c}$ is a dimensionless
voltage, where $V_{c}$ is a characteristic voltage. By means of simple
manipulations, we can rearrange
{Eq. (\ref{eqn2})} as
\begin{eqnarray}
z \ U_{n} = \gamma\  (U_{n+1} + U_{n-1} ) + \epsilon\  \delta_{n 0}\
U_{n}
\label{eq3}
\end{eqnarray}
with
\begin{eqnarray*}
z &\equiv& \Omega^2 - 2 \omega_{1}^2 - \omega_{2}^2\\
\gamma&\equiv& -\omega_{1}^2\\
\epsilon&=& -(\Delta C/C) \Omega^2.
\end{eqnarray*}
where $\omega_{1}^2=1/(L_{1} C)$, and $\omega_{2}^2=1/(L_{2} C)$.

Equation (\ref{eq3}) describes formally a single impurity tight-binding model, whose Hamiltonian is given by
\be
H = H_{0} + H_{d}.\label{four}
\ee
Using Dirac's notation, we can express $H_{0}$ and $H_{1}$ in terms of Wannier functions $\{ |n\rangle \}$ as
\be 
H_{0} = \gamma \sum_{n}(|n+1\rangle \langle n| + |n\rangle \langle n+1|), 
\ee
the (undisturbed) lattice Hamiltonian, and
\be
H_{d} = \epsilon \ |0\rangle \langle 0|,
\ee
the {\it defect} (or impurity) Hamiltonian. 
The Hamiltonian $H_{0}$ describes the lattice without the impurity and, as is well-known, possesses plane-wave eigenvectors
\be 
<n|k> = (1/\sqrt{N}) \exp(i k n)
\ee
and eigenvalues
\be
z_{k} = 2 \gamma \cos(k)
\ee
or, in terms of the electrical lattice parameters, 
\be
\Omega^2 = 4\  \omega_{1}^2 \sin^2(k/2) + \omega_{2}^2
\ee
Thus, the system is able to support the propagation of electrical
waves that form a
band extending (in terms of $\Omega^2$)
from $\omega_{2}^2$ to $\omega_{2}^2 + 4
\omega_{1}^2$.
Notice that in the infinite lattice limit this band would be a
continuous spectrum, while in the finite lattice case, the
boundary
conditions select a specific set of wavenumbers $k$
(e.g. $k_m=m \pi/N$ with $m$ integer running up to $N$ for periodic
boundary conditions), and thus
only the corresponding frequencies are observed.

\section{The Green's function formalism}

Given a Hamiltonian $H$, the lattice Green's function is defined as \cite{green1, green2, green3}
\be
G(z) = 1/(z-H)
\ee
In our case, $H = H_{0} + H_{d}$. Treating $H_{d}$ formally as a perturbation, we can express $G(z)$ as
\be
G(z) = G^{(0)} + G^{(0)}\ H_{d}\  G^{(0)} + G^{(0)}\ H_{d}\ G^{(0)}\ H_{d}\  G^{(0)} +\cdots
\ee
where $G^{(0)}=1/(z-H_{0})$ is the unperturbed Green function. Inserting $H_{d} = \epsilon |0\rangle \langle 0|$, we have
\begin{eqnarray*}
G(z)&=&G^{(0)} + \epsilon\ G^{(0)}|0\rangle \langle 0|G^{(0)} + \epsilon^2 G^{(0)} |0\rangle \langle 0|G^{(0)}|0\rangle \langle 0|G^{(0)} +\cdots\\
&=& G^{(0)}+\epsilon \ G^{(0)} |0\rangle 
\left( \sum_{n=0}^{\infty} \epsilon^{n} G_{0 0}^{(0) n}\right)
\langle 0|G^{(0)}  \\
& = & G^{(0)} + {G^{(0)}|0\rangle \ \epsilon\  \langle 0| G^{(0)} \over{1-\ \epsilon \ G_{0 0}^{(0)}}}.
\end{eqnarray*}

It can thus be easily proven that the eigen-energies of the localized modes are given by the poles of $G$, while the probability amplitudes are given by the residues of $G$ at those poles.

\subsection{Electrical impurity in the bulk}

\begin{figure}[t]
\includegraphics[angle=0, width=1.0\linewidth]{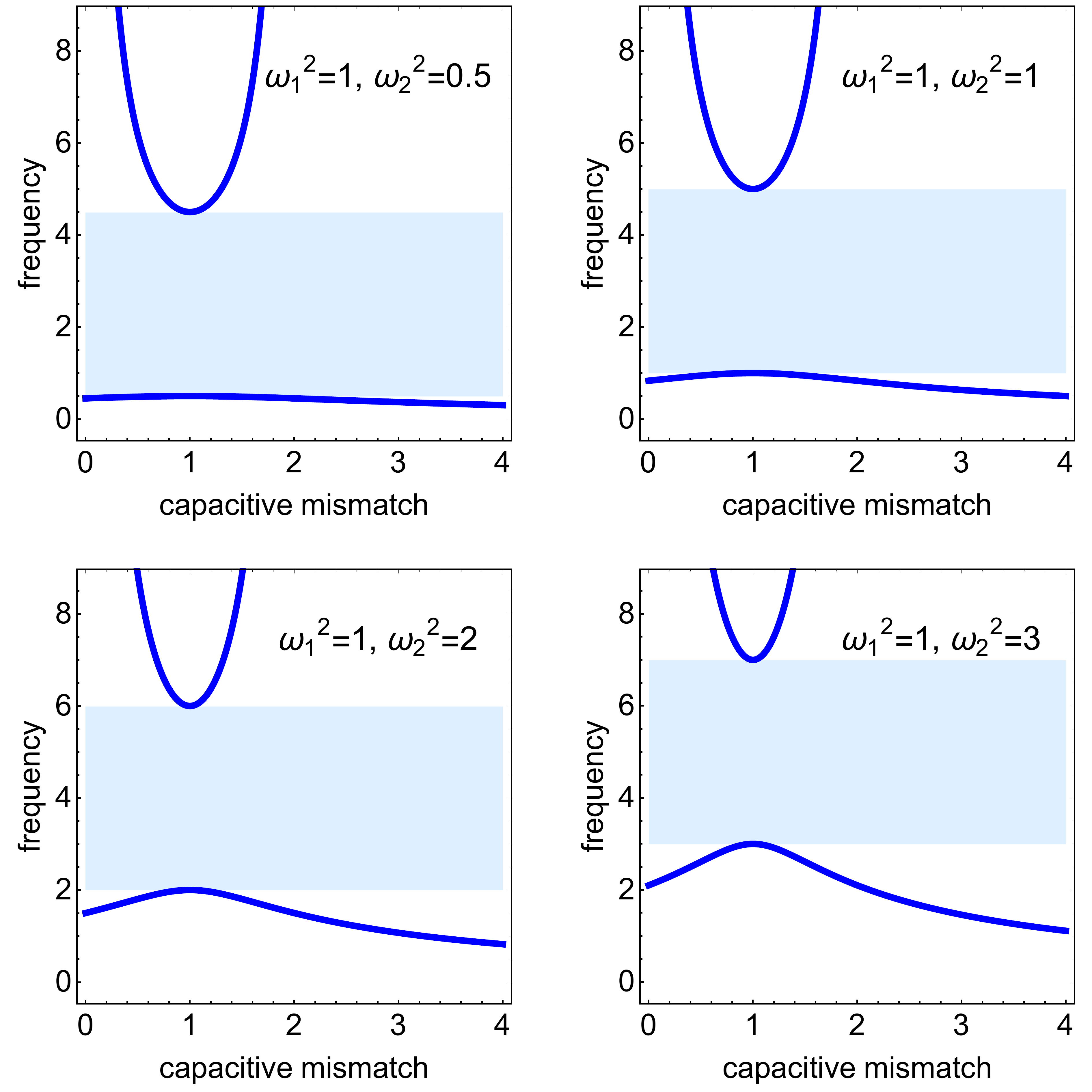}
\caption{\protect
Eigenfrequencies of the localized mode versus the capacitive mismatch, for
several $\omega_{1}, \omega_{2}$ values.
Recall that for $\delta<1$ only the higher frequencies (above the
band)  are physically
relevant,
while
for $\delta>1$ only the lower frequencies (below the band) are physically relevant
(among the two branches shown). This will be an important observation
to bear in mind also
for the physical experiments that will follow.
\label{fig2}}
\end{figure}

Let us consider a capacitive defect that is located far away from the
boundaries of the system. In that case we have from~\cite{green1} that:
\be
G_{0 0}^{(0)}(z) = {\mbox{sgn}(z)\over{\sqrt{z^2-(2 \gamma)^2}}}.
\ee
After solving the energy equation $1/\epsilon = G_{0 0}^{(0)}(z_{b})$, one obtains,
\be z_{b} = \pm \sqrt{\epsilon^2 + (2 \gamma)^2}
\ee
In terms of our electrical parameters, this leads to
\be
\Omega^2={{-(2 \omega_{1}^2+\omega_{2}^2)\pm \sqrt{4 
\omega_{1}^4 +(\delta-1)^2 \omega_{2}^2(4 \omega_{1}^2 +\omega_{2}^2)}}\over{\delta (\delta-2)}}, \label{eq14}
\ee
where $\delta=C_{0}/C$ is the capacitive mismatch.
However, it should be kept in mind that only one of the branches
corresponds to each case: the $(-)$ one in Eq.~(\ref{eq14}) for
$C_0<C$ or $\delta<1$, and the $(+)$
one  in Eq.~(\ref{eq14})
for $C_0>C$ or $\delta>1$.
Figure~\ref{fig2} shows the theoretically predicted energy branches
as a function of the capacitive mismatch $\delta$, for several different resonant frequencies $\omega_{1}^2, \omega_{2}^2$. All eigenfrequencies $\Omega^2$ lie outside of the band. In the absence of the defect, i.e., for $\delta=1$, the solutions touch the band right at the edges of the band. These edge modes are the ones that detach from the band due to the presence of the defect.

For the bound state profile $|b\rangle$, we start from
$|b\rangle = \sum_{n} b_{n} |n\rangle$, where $b_{n}$ is the mode amplitude profile, and is given by the residue of $G(z)$ at $z=z_{b}$. 

\be 
b_{n} = {G_{n0}(z_{b})\over{\sqrt{-G'_{00}(z_{b})}}}
\ee
where
\be
G_{n0}(z) = {\mbox{sgn}(z)\over{\sqrt{z^2-(2 \gamma)^2}}}
\left\{-\left({z\over{2|\gamma|}}\right)+\mbox{sgn}(z) \sqrt{\left({z\over{2 \gamma}} \right)^2 - 1} \right\}^{|n|}.
\ee
We obtain:
\be
b_{n} = \mbox{sgn($z_{b}$)} {(z_{b}^2-(2 \gamma)^2)^{1/4} \over{|z_{b}|^{1/2}}}
\left\{-\left({z_{b}\over{2|\gamma|}}\right)+\mbox{sgn($z_{b}$)} \sqrt{\left({z_{b}\over{2 \gamma}} \right)^2 - 1} \right\}^{|n|}
\ee
where, $z_{b}=\Omega^2-2 \omega_{1}^2-\omega_{2}^2, \ |\gamma|=
\omega_{1}^2$, and $\Omega^2$ is given by Eq.~(\ref{eq14}). Recall that
$z_b>0$ for $\delta<1$, while $z_b<0$ for $\delta>1$.
\begin{figure}[t]
\includegraphics[angle=0, width=1\linewidth]{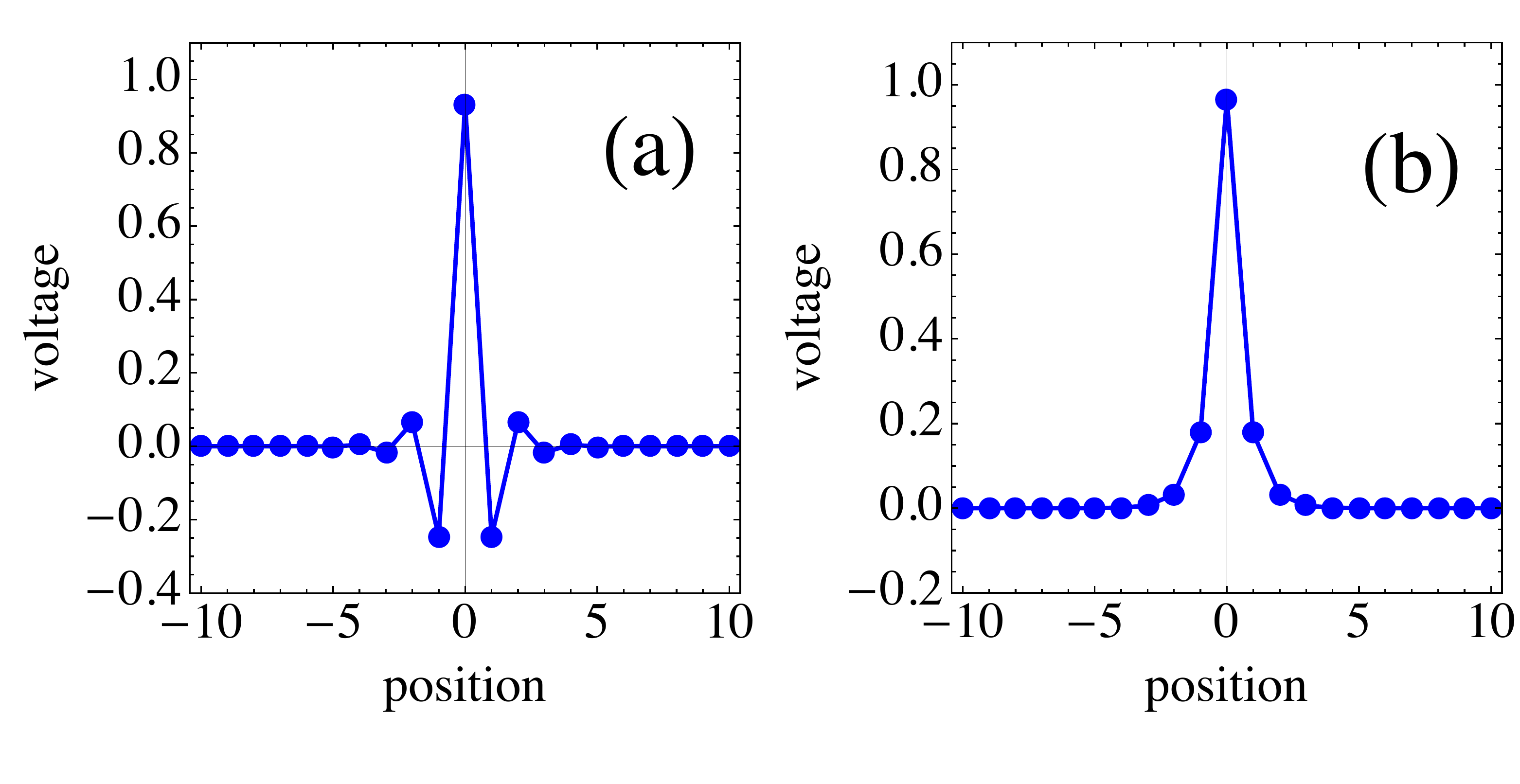}
\caption{\protect
Spatial profiles of the localized electrical bulk modes for
$\omega_{1}=1=\omega_{2}$; (a) $\delta=0.5$ (upper branch), (b)
$\delta=3$ (lower branch).
\label{fig3}}
\end{figure}
Figure~\ref{fig3} shows the spatial profile of the bound state for
different
electrical parameters, namely for $\delta<1$ and $\delta>1$. The mode
width decreases with either an increase in capacitive mismatch
(associated with $\delta$) or an increase in frequency mismatch $|\omega_{1}^2-\omega_{2}^2|$. 

\subsection{Electrical impurity at the boundary}
\begin{figure}[t]
\includegraphics[angle=0, width=0.95\linewidth]{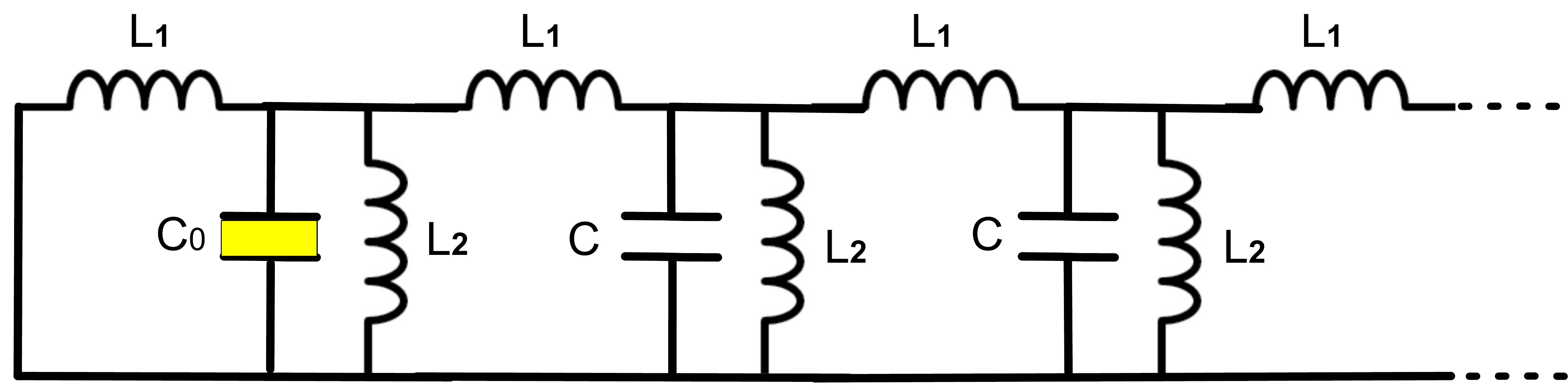}
\caption{\protect
Bi-inductive electrical lattice containing a single surface capacitive
impurity  (in line with Remoissenet et.al.~\cite{remoissenet1, remoissenet2}). 
\label{fig4}}
\end{figure}

Now we consider the case where the defect is placed at the very
surface of a semi-infinite electrical array
(Fig.~\ref{fig4}).
Computation of the proper Green's function for this case requires realizing that, even though the main formalism is still the same as in section A, the unperturbed Green's function $G_{m n}^{(0)}$ needs to be computed by the method of mirror images: since there is no lattice to the left of $n=0$, $G_{m n}^{(0)}$ must vanish identically at $n=-1$. This means $G_{m n}^{(0)}(z)=G_{m n}^{\infty}(z)-G_{m,-n-2}^{\infty}(z)$, where $G_{m n}^{\infty}(z)$ is the unperturbed Green's function of the bulk case: $G_{m n}^{\infty}(z)=\mbox{sgn}(z)(1/\sqrt{z^2-1})[z-\mbox{sgn}(z)\sqrt{z^2-1}]^{|n-m|}$. The procedure for computing the frequency of the surface bound state and its profile is exactly the same as in section III A, and it is detailed in the Appendix.

Figure~\ref{fig5} shows the eigenfrequency $\Omega^2$ of the bound state as a function of the capacitive mismatch $\delta$.
\begin{figure}[t]
\includegraphics[angle=0, width=0.8\linewidth]{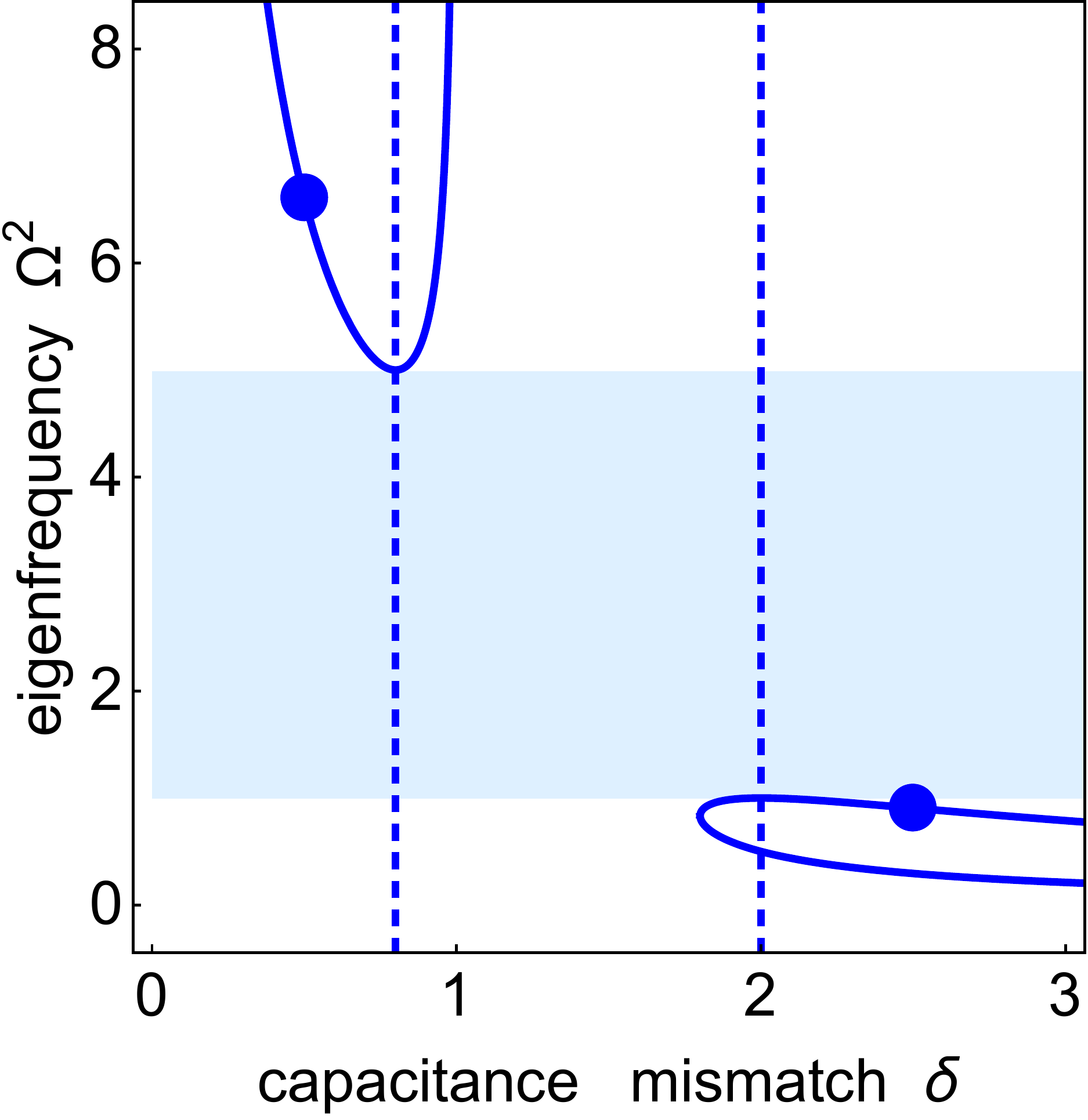}
\caption{\protect
Eigenfrequencies $\Omega^2$ of the electrical surface mode, in terms
of the capacitance mismatch $\delta$, for
$\omega_{1}^2=1=\omega_{2}^2$. The dashed lines mark the position of the critical 
$\delta$ values $\delta_{2}=0.8$ and $\delta_{3}=2$. 
\label{fig5}}
\end{figure}
An interesting feature of $\Omega^2$
in this case (with the boundary) concerns the  forbidden bands  inside which
there are no bound states.
In terms of frequencies, there
exists the well-known frequency band that extends from $\omega_{2}^2$
up to $\omega_{2}^2 + 4 \omega_{1}^2$. In terms of the parameter
$\delta$, there exists 
an interval in capacitive mismatch that extends from $\delta=1$ to
\be 
\delta_1 = {( 2\ \omega_{1}^2+ \omega_{2}^2 )^2\over{4 \omega_{1}^2
    \omega_{2}^2 + \omega_{2}^4}},
\label{deltac}
\ee
inside which the mode is complex. The most important intervals
however, are the ones originating from the condition that the
capacitance mismatch $|\delta-1|$ be sufficiently large to produce a
bound state. These are given by $\delta < \delta_2$ and $\delta > \delta_3$, where
\be
\delta_2 = {3 \omega_1^2+\omega_2^2\over{4 \omega_1^2+\omega_2^2}} < 1,\ \ \delta_3 = 
1+\left( {\omega_1\over{\omega_2}} \right)^2 >1.
\ee
The origin of this condition lies in  the fact that the real part of the surface Green's function, $G_{0 0}^{(0)}(z)$ for $z$ outside the band, is bounded from above and below, unlike the case of the  bulk Green's function.
Out of the different
critical $\delta$ values only $\delta_2$ and $\delta_3$ are relevant since
$\delta_2 <1$ and $\delta_3 > \delta_1$. It can be also proven that at
$\delta=\delta_{2,3}$ the energy curves touch the linear band. These
features can be appreciated in
Fig.~\ref{fig5}.
\begin{figure}[h]
\includegraphics[angle=0, width=1\linewidth]{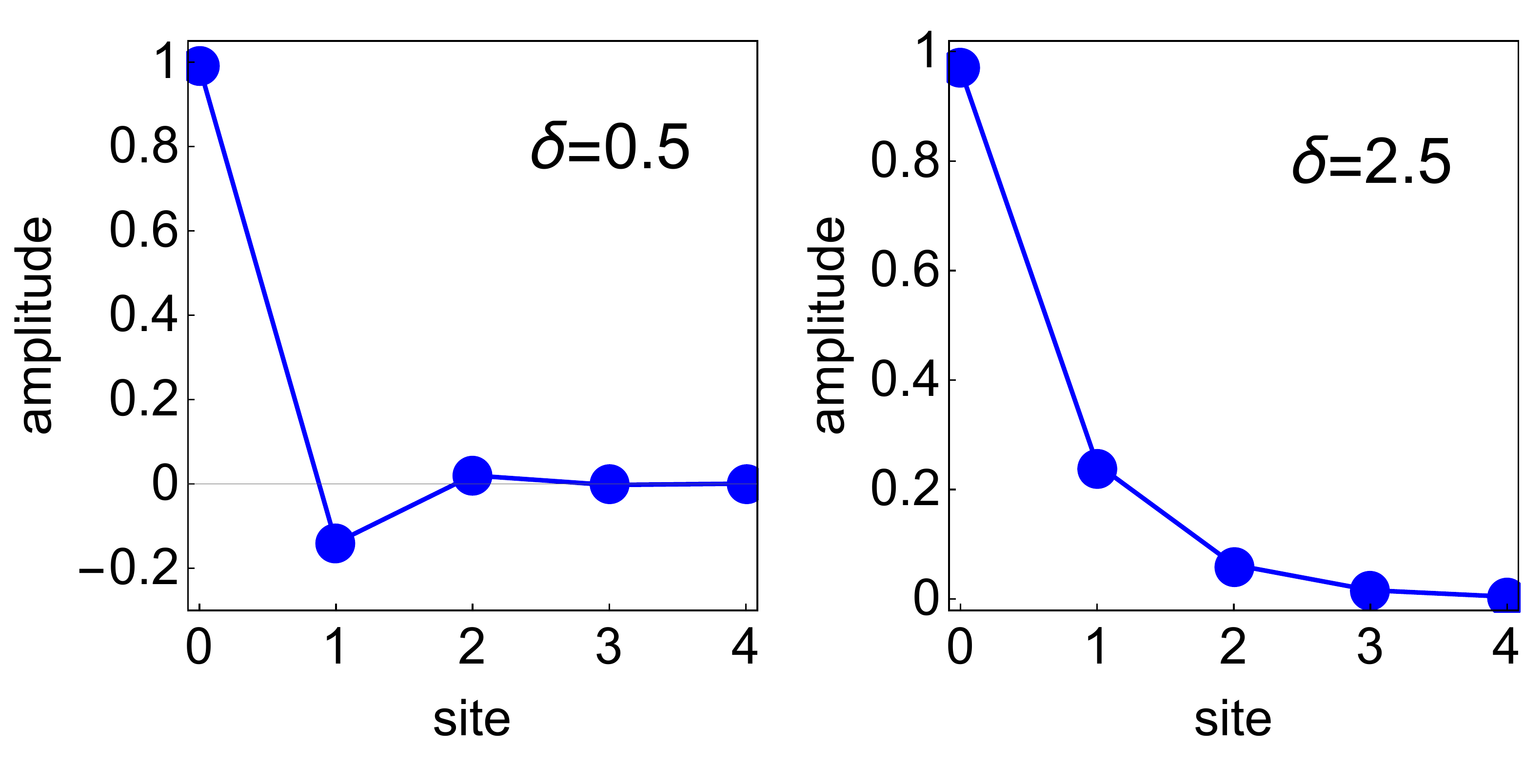}
\caption{\protect
  Spatial profiles of  the localized surface modes marked in
  Fig.~\ref{fig6}
  for $\delta=0.5$ and $\delta=2.5$.
\label{fig6}}
\end{figure}
Figure~\ref{fig6} shows once again
some spatial mode profiles. We now turn to the experimental
realization of
the above localized modes.

\section{The experiment}
We have built the bi-inductance lattice shown in Fig.~\ref{fig1} made
up of discrete electrical elements of inductors and capacitors. The
inductors were ferrite (radial-lead) inductors of 470 $\mu$H
inductance and about 1 $\Omega$ resistance (ESR - equivalent series
resistance). We used these inductors both as the $L_1$ and $L_2$
elements in a lattice with $\omega_2^2 \approx \omega_1^2$. {The
  inductance values were all within 1 percent of one another}. One set
of measurements was also done on a lattice with $\omega_2^2 \approx 2
\omega_1^2$, and there we also incorporated 975 $\mu$H inductors with
an ESR of 2 $\Omega$. The fixed capacitors, $C$, were all of 1 nF
capacitance,
{to within 2 percent of one another}. In addition, in order to excite any mode we incorporated a driving signal using an arbitrary function generator (Agilent 33220A) whose signal was injected into the electrical lattice via 10 k$\Omega$ resistors. The voltages at each lattice node were monitored using multi-channel DAQ boards (National Instruments PXI-6133).

The impurity was introduced via the replacement at one lattice site of the 1 nF capacitor with a varactor diode (NTE-618) whose capacitance can be continuously varied by applying a bias voltage. The capacitance-voltage relationship for the varactor diode was carefully mapped out by separately measuring the resonance curve of an RLC circuit that incorporated this diode in parallel with a known inductor. When introducing such a bias voltage across the varactor diode, care must be taken, however, to ensure that none of the inductors in the lattice experiences a DC potential drop across them. This was achieved by using a large (2 $\mu$F) block capacitor in series with the varactor diode. 
In essence, then, we can continuously vary the $\delta$-parameter with a DC power supply that controls the bias voltage across the varactor diode. This yields a capacitance range for $C_0$ from about 900 pF down to about 150 pF. Given our choice of $C=1$ nF, in order to attain the $\delta>1$ regime, we can place this varactor diode in parallel with a fixed 1 nF- or 2 nF capacitor.     

\subsection{Impurity in the Lattice Bulk}
      
\begin{figure}[h]
\includegraphics[width=0.975\linewidth]{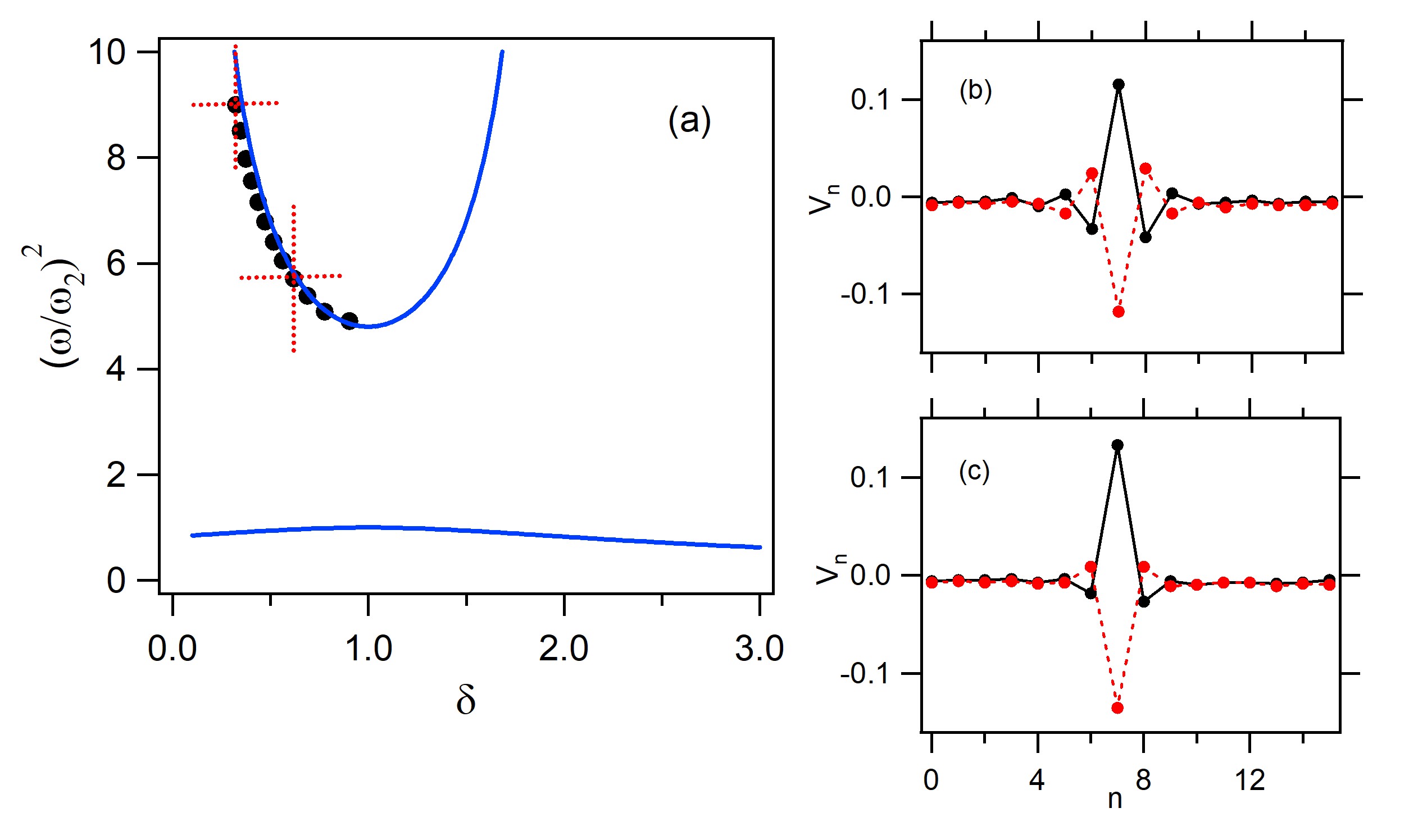}
\caption{\protect (a) The experimental frequencies of the upper-branch
  impurity modes as a function of capacitance-mismatch parameter
  $\delta$. The data fits the theoretical prediction quite well. (b)
  and (c) depict measured mode profiles at $\delta=0.62$ and
  $\delta=0.32$, respectively. The times at which the center node is
  most positive and most negative are shown, respectively, by solid
  and dashed lines.
\label{exp1}}
\end{figure}

To compare the experimental observations to the theoretical
predictions in the previous section, Fig.~\ref{exp1}(a) depicts the
square of the normalized frequencies, $(\omega/\omega_2)^2$ as a
function of the capacitance-mismatch parameter, $\delta$, {which measures the strength of the impurity,} for the upper-branch
localized modes. We see that the experimental data points (black
circles) match the theoretical curve very well for the parameter
values of $\omega_2^2=1$ and $\omega_1^2=0.95$. 
Importantly, as we highlighted in the theoretical section only one of
the two solutions with frequencies above the band is selected
in line with the theoretical prediction.
To excite these impurity modes, we can use a sinusoidal waveform that is swept through a range of frequencies while simultaneously recording the impurity mode response. In this way a spectrum is generated, and the mode frequency is obtained from the peak's center frequency within that spectrum.  The driving waveform can be applied uniformly at each site of the lattice, or alternatively only at the impurity site. Both driving methods yield the same frequency information, although the exact mode profile does depend slightly on the driving method. Here we show results corresponding to local driving. 

From the inductor values themselves, i.e., $L_1=L_2=470 \mu$H, we would expect $\omega_1^2=1$. The fact that this value is slightly lower in the experimental system, however, is confirmed independently by separately measuring the zone-boundary (ZB) mode frequency. Here a spatially staggered driver is employed to probe the resonance curve associated with the ZB mode, and its center frequency is recorded at around 502 kHz. Given that the zone-center mode is at $\omega_2/(2\pi) =$ 228 kHz, we would expect the ZB to be at approximately 510 kHz. The fact that it is found at somewhat lower frequency indicates that the effective $\omega_1$ is also lower by a small amount.      

Figure panels \ref{exp1}(b) and (c) show the measured spatial profiles of two impurity modes, corresponding to $\delta=0.62$ and $\delta=0.32$, respectively. The node voltages measured at two particular times are shown at which the voltage at the center node (n=7) is most positive and most negative, respectively. As also seen in the theoretical waveforms, the farther $\delta$ departs from 1, and thus the higher the frequency, the more localized the impurity mode is in space. 

\begin{figure}[h]
\includegraphics[width=0.85\linewidth]{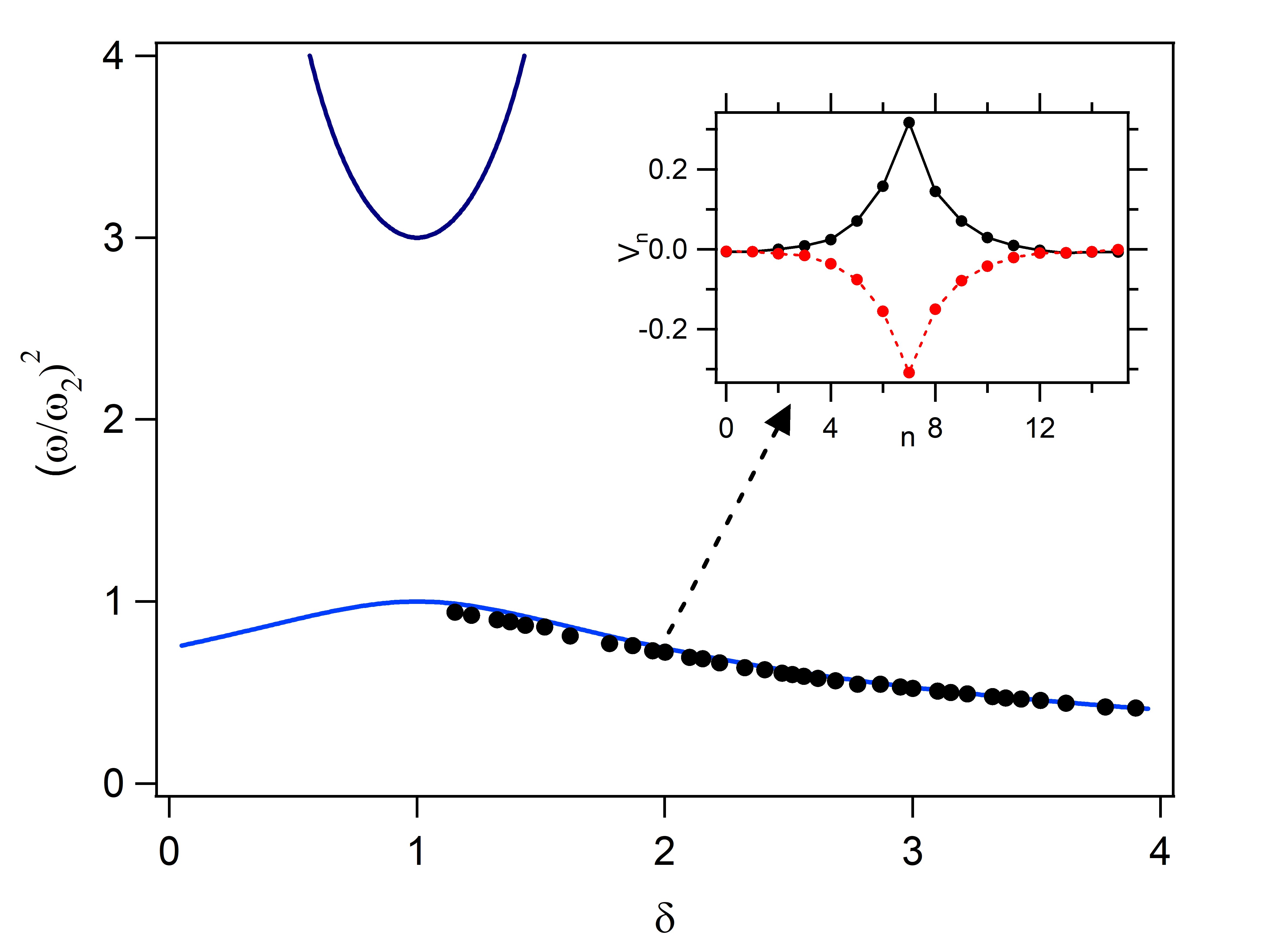}
\caption{\protect The experimental frequencies of the lower-branch
  impurity modes as a function of the mismatch parameter
  $\delta$. Again, the data fits the theoretical prediction quite
  well, with some slight deviation near $\delta=1$. As explained in
  the previous section only the portion of the branch with $\delta>1$
  is physically relevant and hence experimentally observed.
 The inset depicts the measured mode profile at $\delta=2.0$.
\label{exp2}}
\end{figure}
We now turn to the lower-branch modes in this lattice.
From the theoretical picture of Fig.~\ref{fig2}(b) and (c), we discern that when $\omega_2$ is raised relative to $\omega_1$, the branch descends away from the bottom of the plane-wave spectrum more rapidly. (The same is also true in the case when $\omega_1$ is lowered relative to $\omega_2$.) For this reason, we decided to decrease the coupling strength - and thus $\omega_1$ - while leaving $\omega_2$ unchanged when observing the lower branch. This was accomplished by using coupling inductors of $L_1=975 \mu$H.  

Figure \ref{exp2} depicts the experimental results on the lower
branch. The lines represent the theoretical curves given in
Eq.~(\ref{eq14}), for $\omega_2^2=1$ and $\omega_1^2=0.475$. We again see
good agreement between theory and experiment, although some deviation
is observed for $\delta$-values close to 1, where the localized mode
becomes quite broad (and where perhaps a larger lattice would thus be
needed). 
Recall that per our theoretical analysis, only the portion of the
branch
with $\delta>1$ is physical (and thus experimentally traceable).
The inset panel shows the impurity mode profile. We see that contrary to the upper-branch modes where neighboring sites oscillate out-of-phase, for these lower-branch impurity modes the nodes all oscillate in-phase. Thus, the impurity modes in both instances share the basic symmetry with that plane-wave mode into which they merge as $\delta \rightarrow 1$. 

\subsection{Impurity at the boundary}
It is straightforward to transform the electrical lattice used so far into a lattice where the impurity is at its boundary. We simply disconnect the far end of the coupling inductor at n=7 and instead ground it. We also return to the case $L_1=470 \mu$H, such that $\omega_2^2=1, \omega_1^2=0.95$.   
\begin{figure}[h]
\includegraphics[width=0.85\linewidth]{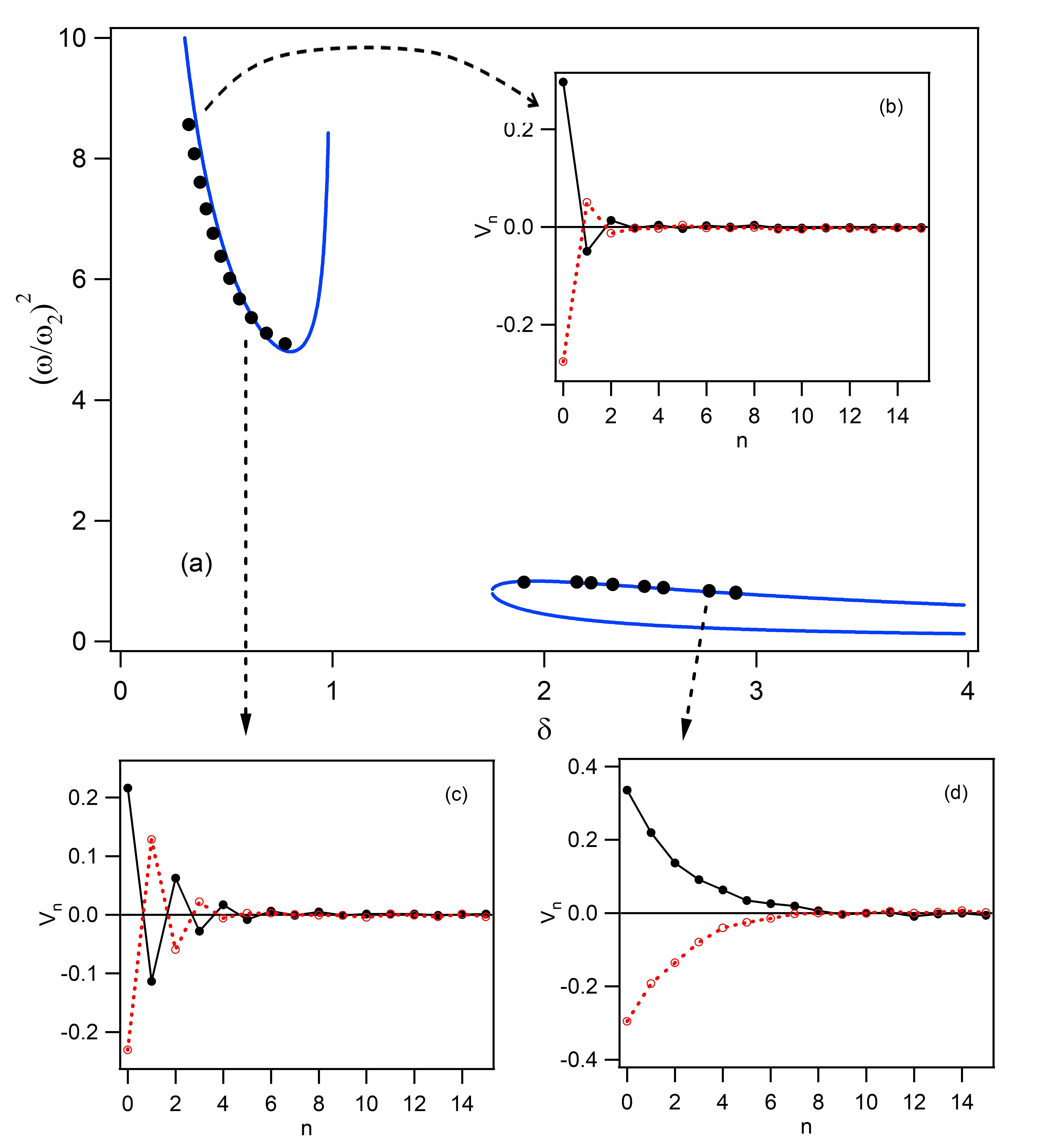}
\caption{\protect (a) The experimental frequencies of the surface
  impurity modes as a function of the capacitance-mismatch parameter $\delta$;
  both upper-branch and lower-branch solutions are shown. The
  theoretical prediction is depicted by the solid curves with only the
  left branch being relevant for $\delta<\delta_2$ and the top branch for
$\delta > \delta_3$. (b)-(d) show the measured profiles of three particular surface modes, located at $\delta=0.32$, $\delta=0.62$, and $\delta=2.78$, respectively.
\label{exp3}}
\end{figure}

Confirming the theoretical prediction, no surface modes are found
within the interval of $0.80<\delta<1.95$. For $\delta<\delta_2=0.80$, we observe upper-branch solutions, and for $\delta>\delta_3=1.95$ lower-branch solutions. Their frequencies nicely match the theoretical prediction, as seen in Fig.~\ref{exp3}(a).
Figure panels~\ref{exp3}(b) and (c) show the spatial profiles of the
upper-branch modes measured at the two $\delta$-values of 0.32 and
0.62, respectively. As expected, we clearly observe the mode widening
in space as $\delta$ approaches 1. Panel (d) depicts the lower-branch
mode at $\delta=2.78$. The width of the mode is somewhat larger than
what is obtained in the analysis; one possible reason might be that
experimentally, some part of the uniform mode may still be excited.

\subsection{Quench experiments}   

In order to examine the impact of quenches and the response of the system between the two
fundamentally different regimes examined ($\delta<1$ and $\delta >1$
in the bulk, and correspondingly also in the case of the surface), we 
devise the following quenching experiment. 
We use a single pole double throw (SPDT) analog switch, the chip
ADG436, to rapidly switch between two capacitor values for $C_0$: 0.57
nF and 1.43 nF. These two values were chosen because they are
symmetrically situated relative to 1 nF.
Again, the driving of the lattice can be either local (at the impurity) or the spatially homogenous (shown
in the following figures). 
For the upper-branch case in Fig.~\ref{exp4}(a), the driving frequency
corresponded to $\Omega^2=5.52$ (equivalent to 536 kHz). We see that at this frequency and for
$C_0=0.57$ nF (corresponding to a $\delta$ of 0.57), the upper-branch
impurity mode can be stably excited. When we abruptly switch to
$\delta=1.43$ at time $t=200 \mu$s, however, the mode disintegrates
and does not reconstitute itself.  That is to say, there is no
corresponding
physical branch on the other side at $\delta=1.43$.

\begin{figure}[h]
\includegraphics[width=0.85\linewidth]{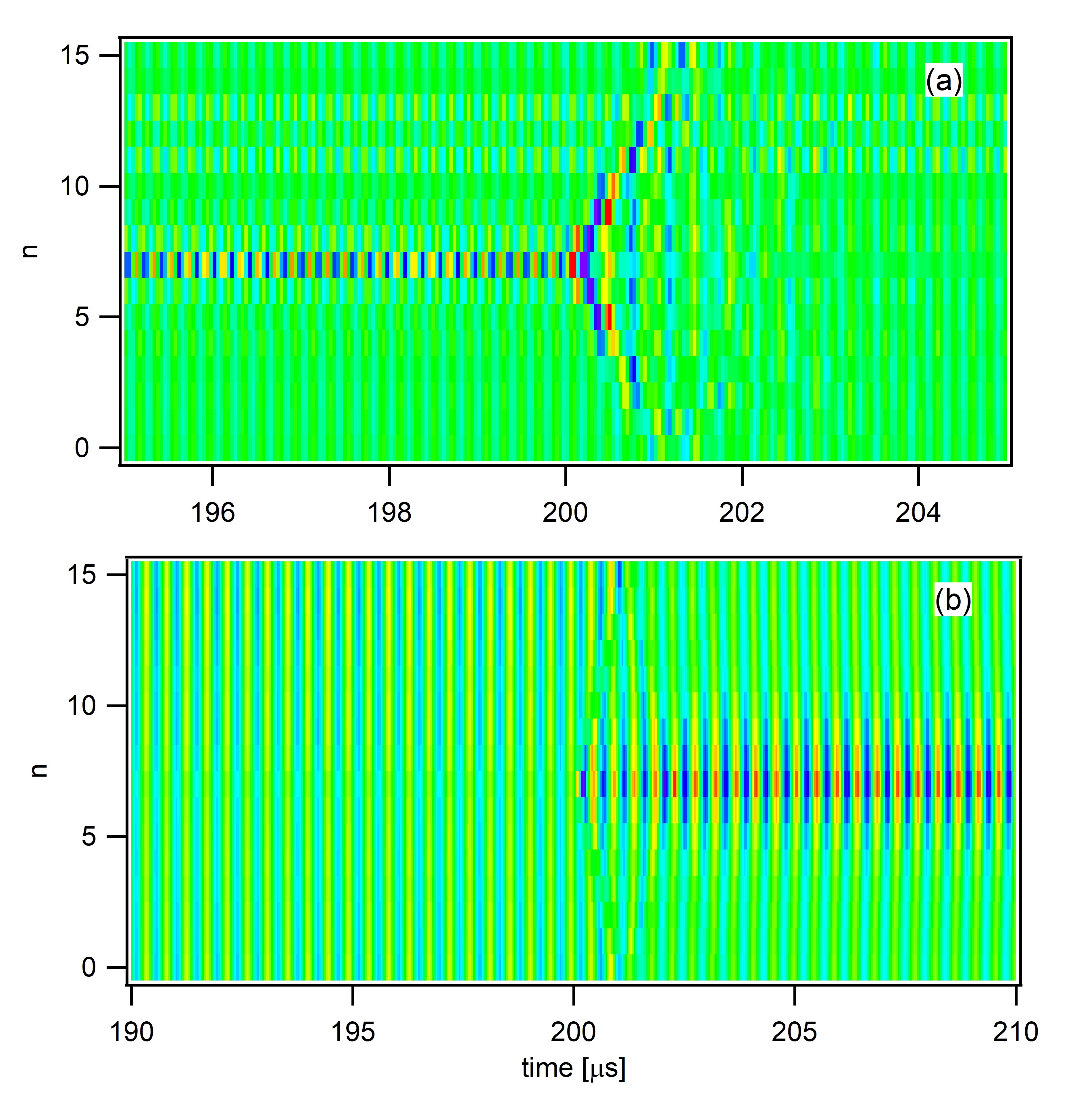}
\caption{\protect Space $n$-time $t$ contour plot evolution of quench
  experiments for the voltage $V_n(t)$: (a) from $\delta=0.57$, we
  quench to $\delta=1.43$ in the case of the upper branch; (b) the
  same
is done in the case of the lower branch. The two panels clearly
illustrate in line also with the theory that the upper frequency
branch only
exists for $\delta<1$, while the lower frequency branch only for $\delta>1$.
\label{exp4}}
\end{figure}
A similar phenomenon can be discerned in Fig.~\ref{exp4}(b) for the
lower branch, where we chose a frequency of 219 kHz (or $\Omega^2=0.92$). Here no mode
exists at $\delta=0.57$, but upon abruptly switching to $\delta=1.43$, 
the lower-branch impurity mode emerges. We should mention that if we
start with the larger value of 
$\delta$ and then switch to the lower one, no impurity mode is
generated in that case either.

\begin{figure}[h]
\includegraphics[width=1.05\linewidth]{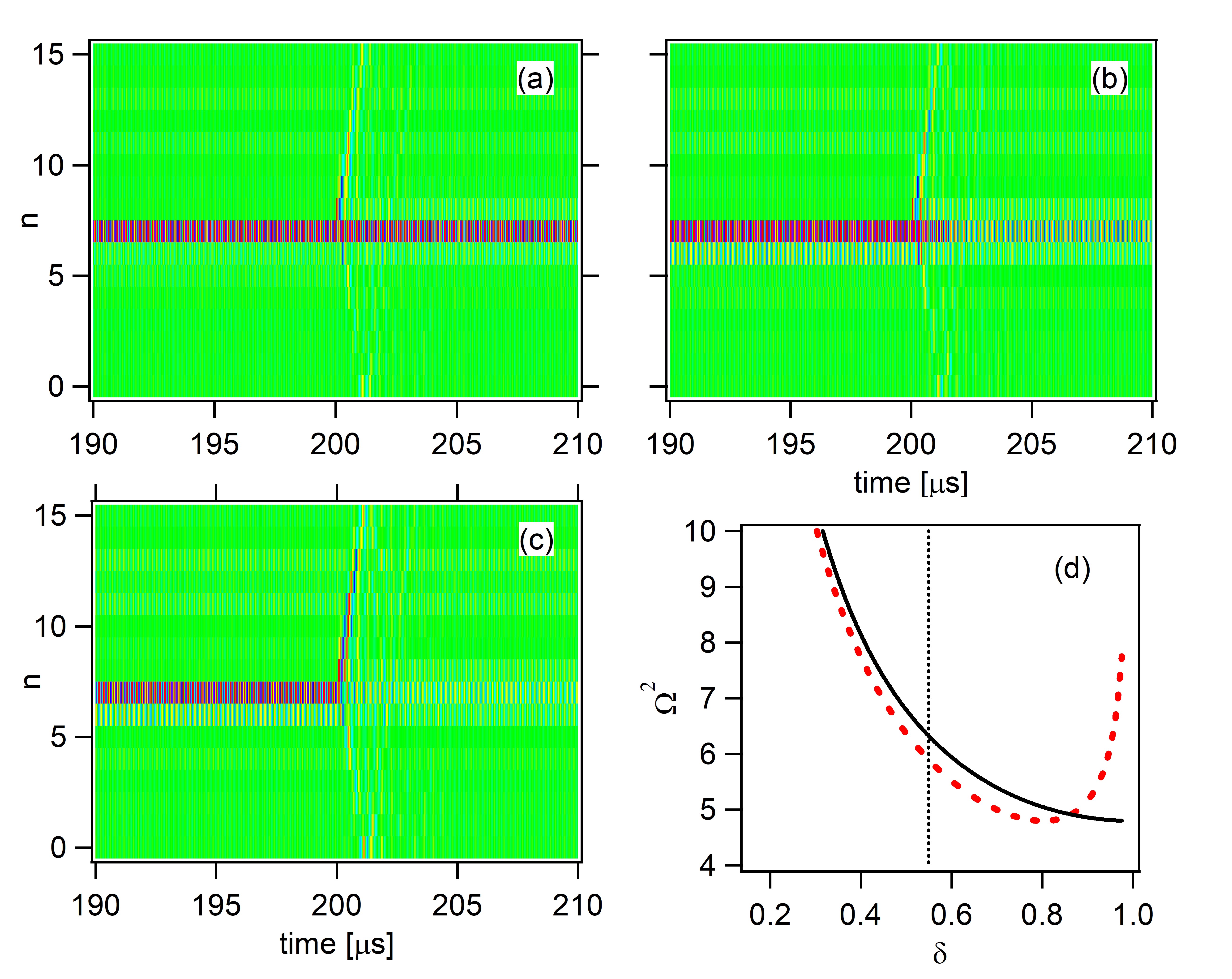}
\caption{\protect {Switching between a surface mode and a bulk mode by closing a switch to complete the ring at $t=200 \mu$s. The driver frequency corresponds to an $\Omega^2$ of (a) 6.34 (b) 5.86 and (c) 5.65. While the surface impurity mode can be generated at all three frequencies, we see the bulk impurity mode unable to sustain itself in (c). This makes sense when comparing the frequencies of the solution branches in (d), where the dotted vertical line is the $\delta$ chosen in these experiments, and the solid and dashed lines are the upper-branch bulk and surface modes, respectively.} 
\label{exp5}}
\end{figure}

Another appealing switching experiment that can be performed in the
present setting involves
transforming the impurity mode from 
a surface one to a bulk one
by
using the analog switch to 
temporarily impose a boundary in the lattice that is otherwise
ring-like. {The switch connects the far end of the $L_1$ inductor at $n=7$ first to ground and then abruptly to the $L_1$ inductor at $n=8$.}
Here we chose a $\delta$ of 0.47 and excited the lattice with a
spatially homogenous driver at various upper-branch frequencies. This
is illustrated in Fig.~\ref{exp5}(a), where the frequency corresponds to $\Omega^2=6.34$. 
The switch from boundary lattice bearing a surface to a ring
lattice (i.e., a bulk one with periodic boundary conditions) occurs at 200
$\mu$s. We see that the initial surface mode naturally
 transforms itself into the bulk mode after the switch. Figure
 \ref{exp5}(b) and (c) decrease the driver frequency consecutively, to
 $\Omega^2=5.86$ and 5.65, respectively. It is clear that the bulk mode is
 weakened in (b) and then disappears altogether in (c), while the
 surface mode is still generated at these frequencies. 

{This last result makes sense upon close comparison of
  Figs.~\ref{exp1}(a)  and~\ref{exp3}(a), which (for convenience) is
  shown in Fig.~\ref{exp5}(d). When superimposing the solution
  branches in the two cases, we see that the upper-branch surface
  modes have lower frequency compared to the bulk modes for all
  allowed $\delta$-values
  of the former (i.e., for $\delta < \delta_2$). Since the
  experimental system features dissipation and thus potential
  relaxation to
  the modes, those modes have some width in frequency
  and therefore can overlap, such that at a given driving frequency
  both can be
  excited. However, if the driving frequency is repeatedly lowered, we
  first lose the
  bulk mode and only later the surface mode.}

\vspace{4cm}
\section{Conclusions \& Future Challenges}
We have studied both theoretically and experimentally an electrical
{transmission-line lattice} possessing a capacitive impurity {point-}defect located either at the
bulk or at the boundary of the
circuit. By using the formalism of lattice Green's functions, we were
able to predict the existence of localized modes for which the voltage
decreases in space as we move away from the defect position. We obtained
in closed form the bound-state 
frequency  of the impurity state, as well as the bound-state profile as
a function of the parameters of the
electrical circuit.
Very good  agreement was observed between the theoretical predictions
and the corresponding experimental results both for the bulk and also
for the surface impurity modes. Additionally, quench-type experiments 
were performed, by rapidly switching a lattice parameter
and observing how the voltage profile adapts itself to such
a modification. Also an interesting scenario of a switch from a
bulk lattice to a surface one was explored and the spontaneous transformation of the
bulk modes to surface ones was elucidated.
{Such experiments were found to be in accord with the analytical
  picture
  characterized by its relevant solution branches.} 


Naturally, this study and the systematic benchmarking of the linear
lattice properties paves the way towards further explorations. At the
linear level, the theory of Green's functions also permits an
analytical characterization of the transmission problem from an
impurity as a function of the wavenumber of the incoming wave, as well
as of the impurity (and lattice) parameters. This is a technically
more challenging problem as it arguably requires larger lattices than the ones
considered herein, yet is, in principle, experimentally tractable. On the other hand,
numerous studies have focused on nonlinear impurities in a linear
lattice~\cite{tsir1,moli,ourmario} as the first among the many
different possible scenarios involving nonlinearity in the present setting. This
is a natural next step for our studies, and will be considered in
future publications.  

\vspace{2cm}

\acknowledgments
MIM acknowledges support from Fondo Nacional de Desarrollo
Cient\'{\i}fico y Tecnol\'{o}gico (Fondecyt) Grant 1160177. PGK
gratefully acknowledges support from NSF-DMS-1809074, the
hospitality of the Mathematics Institute of the University of Oxford
and the support of the Leverhulme Trust.

\section{APPENDIX: Surface modes}

The bound state energies are given by the poles of $G_{n,m}$,   that is, $G_{0,0}^{(0)}(z) = 1/\epsilon$,
where \ \ \ $G_{n,m}^{(0)} = G_{n,m}^{\infty} - G_{n,-m-2}^{\infty}$, \ \ \  
$\epsilon = (1-\delta)\Omega^{2}/2\omega_{1}^{2}$\ \ \   and\  \ \ $z = \Omega^2 - 2 \omega_1^2-\omega_2^2$.

We obtain the eigenfrequencies as a function of the capacitance mismatch $\delta$, as the residues of $G_{n,m}$ at the poles:
\be
\Omega_{\pm}^2 = {(2 \omega_{1}^2 + \omega_{2}^2) \pm
 {\sqrt{ \frac{-4 \omega_{1}^4}{\delta-1}+4 \omega_1^2 \omega_2^2 +\omega_2^4}}\over{2 \delta}}.
\ee

The bound state amplitudes are formally given by the residues of the surface Green function $G_{n m}(z)$, that is
\be
b_{n}^{\pm} = {G_{n,0}^{(0)}(z(\delta))\over{\sqrt{-G_{n 0}^{(0)'}(z(\delta))}}}
\ee
where
\begin{align}
  G_{n,0}^{(0)}(z)={\mbox{sgn}(z)\over{\sqrt{z^2-1}}}[\ (z - \mbox{sgn}(z) \sqrt{z^2-1})^{|n|} \\
  - (z - \mbox{sgn}(z) \sqrt{z^2-1})^{|n+2|}\ ]
\end{align}

and 

\be
G_{n,0}^{(0)'}(z)=-{2\ \mbox{sgn}(z)\over{\sqrt{z^2-1}}} (z - \mbox{sgn}(z) \sqrt{z^2-1}).
\ee

\end{document}